\documentclass[fleqn,twoside]{article}
\usepackage{Gc}
\usepackage{amssymb}
\usepackage{amsmath}
\usepackage{graphicx}

\heads{A.B. Balakin and A.E. Zayats}
      {Ray optics in the field of non-minimal Dirac monopole}

\begin{document}
\onecolumn \Arthead{13}{2007}{? (?)}{1}{8}

\Title{RAY OPTICS IN THE FIELD OF NON-MINIMAL DIRAC MONOPOLE}

\Author{A.B. Balakin\foom 1 and A.E. Zayats\foom 2}
{Department  of General Relativity and Gravitation,\\
Kazan University, Kremlevskaya str. 18, Kazan 420008, Russia}

\Abstract {Based on the analogy with non-minimal $SU(2)$ symmetric
Wu-Yang mono\-pole with regular metric, the solution describing a
non-minimal $U(1)$ symmetric Dirac monopole is obtained. In order to
take into account the curvature coupling of gravitational and
electromagnetic fields, we reconstruct the effective metrics of two
types: the so-called associated and optical metrics. The optical
metrics display explicitly that the effect of birefringence induced
by curvature takes place in the vicinity of the non-minimal Dirac
monopole; these optical metrics are studied analytically and
numerically. }

\email 1 {Alexander.Balakin@ksu.ru}%
\email 2 {Alexei.Zayats@ksu.ru}%

\section{Introduction}

The Dirac monopole as a specific static spherically symmetric
solution to the minimal Einstein-Maxwell equations has became a
subject of discussion in tens papers, reviews and books (see, e.g.,
\cite{Dirac,000,00,0,1,3,2}). The motion of massive and massless
particles, which possess electric charge or are uncharged, is
studied in detail (see, e.g., \cite{4,5} and references therein). In
the paper \cite{BaZa07} we introduced and discussed the $SU(2)$
symmetric Wu-Yang monopole of the new type, namely, the non-minimal
monopole with regular metric. Since the non-minimal Wu-Yang monopole
is effectively Abe\-lian, it is naturally to consider the
corresponding analog of that solution in the framework of
non-minimal electro\-dynamics. Mention that non-minimal models with
mag\-netic charge have been discussed earlier (see, e.g.,
\cite{Horn,MHS}), but the exact analytical regular solution obtained
here as direct reduction to the $U(1)$ symmetry is the new one.

The second novelty of the presented paper is the investigation of
photon dynamics in the vicinity of the Dirac monopole accounting the
curvature coupling of the gravitational and electromagnetic fields.
In the presence of non-minimal interaction (induced by curva\-ture)
the master equations for electromagnetic and gravitational fields in
vacuum can be rewritten as the master equations in some effective
anisotropic (quasi)me\-dium \cite{B1,BL05}. This means that two
effective (optical) metrics can be introduced
\cite{HehlObukhov,Perlick,BZ05}, so that the photon propagation in
vacuum interacting with curvature is equivalent to the photon motion
in the effective space-time with the first or second optical metric,
depending on the photon polarization. Even if the real space-time
has a regular metric, the optical metrics can be singular, admitting
the interpretation in terms of the so-called ``trapped surfaces''
and ``inaccessible zones'' \cite{AG1,AG2}. We discuss this problem
in Sect. 3. Numerical modeling of the photon orbits, presented in
Sect. 4, supplement our conclusions.

\section{Master equations and background fields}

We consider the Lagrangian\footnote{Hereafter we use the units
$c=G=\hbar=1$.}
\begin{equation}
{\cal L} = \frac{R}{8\pi}+\frac{1}{2}F_{ik}F^{ik}+\frac{1}{2}{\cal
R}^{ikmn}F_{ik}F_{mn} \label{1}
\end{equation}
to describe the background gravitational and magnetic fields in the
framework of non-minimal Einstein-Maxwell model with the non-minimal
susceptibility tensor
\begin{align}
{\cal R}^{ikmn}=\frac{q}{2} \biggl[
R\left(g^{in}g^{km}-g^{im}g^{kn}\right) - 12 R^{ikmn} + 4
\left(R^{im}g^{kn}+R^{kn}g^{im}-R^{in}g^{km}-R^{km}g^{in}\right)\biggr]\label{2}
\end{align}
linear in curvature. As usual, $R^{ikmn}$ is the Riemann tensor,
$R^{mn}$ is the Ricci tensor, $R$ is the Ricci scalar, $F_{mn}$ is
the Maxwell tensor. This non-minimal susceptibility tensor can be
obtained from the general one (see \cite{BL05}), when the coupling
constants $q_1$, $q_2$, $q_3$ are chosen as follows $q_1=-q<0$,
$q_2=4q$, $q_3=-6q$. The ansatz for the space-time metric is
\begin{equation}
ds^2_{(0)}=N(r)dt^2-\frac{dr^2}{N(r)}-r^2 \left(d \theta^2 +
\sin^2\theta d\varphi^2 \right)\,. \label{3}
\end{equation}
Mention that the choice $g_{tt}g_{rr}=-1$ in our ansatz is supported
by the result, obtained in \cite{BaZa07} for the given relations
between the coupling constants $q_1$, $q_2$, $q_3$. The equations of
electrodynamics \cite{HehlObukhov,EM}
\begin{equation}
\nabla_k H^{ik} =0 \,, \quad \nabla_k F^{*ik} = 0 \label{4}
\end{equation}
with constitutive equations
\begin{align}
H^{ik} = C^{ikmn} F_{mn} \,,\quad C^{ikmn} \equiv \frac{1}{2} \left(
g^{im}g^{kn} - g^{in}g^{km}\right) + {\cal R}^{ikmn} \label{40}
\end{align}
are associated with the Lagrangian (\ref{1}), where $F^{*ik}$ is the
dual Maxwell tensor and $H^{ik}$ is the induction tensor. The
equations (\ref{4}) with (\ref{40}) are satisfied identically, when
the potential of static spherically symmetric electromagnetic  field
out of the point-like magnetic charge $\mu$ is of the form
\begin{equation}
A_k  = \delta_k^{\varphi} A_{\varphi} = - \delta_k^{\varphi} \ \mu
(1-\cos{\theta}) \,. \label{5}
\end{equation}
The corresponding strength field tensor has only one non-vanishing
component
\begin{equation}
F_{\theta \varphi} = - \mu \sin{\theta} \,, \label{6}
\end{equation}
which does not depend on the non-minimal coupling parameter $q$.
Thus, the well-known solution with the magnetic field of the
monopole type
\begin{equation}
B^i \equiv F^{*ik}U_k = \delta^i_r \frac{\mu \sqrt{N}}{r^2} \,,
\quad B(r) \equiv \sqrt{-B^iB_i} = \frac{\mu}{r^2} \,, \label{7}
\end{equation}
satisfies the {\it non-minimal} Maxwell equations (\ref{4}),
(\ref{40}), (\ref{2}). The equations for the gravity field
\begin{equation}
R_{ik} - \frac{1}{2} R \ g_{ik} = 8\pi T^{({\rm eff})}_{ik} \,,
\label{standardform}
\end{equation}
obtained by direct variation procedure from the Lagrangian (\ref{1})
are non-minimally extended, since the effective stress-energy tensor
$T^{({\rm eff})}_{ik}$ has the form
\begin{equation}
T^{({\rm eff})}_{ik} = T^{(0)}_{ik} - q \left[T^{(1)}_{ik} - 4
T^{(2)}_{ik} + 6 T^{(3)}_{ik} \right] \,. \label{effect}
\end{equation}
The quantities $T^{(0)}_{ik}$, $T^{(1)}_{ik}$, $T^{(2)}_{ik}$ and
$T^{(3)}_{ik}$ are given by
\begin{equation}
T^{(0)}_{ik} =  \frac{1}{4} g_{ik} F_{mn}F^{mn} - F_{im}F_k^{ \ m}
\,, \label{part0}
\end{equation}
\begin{align}
T^{(1)}_{ik} = R \ T^{(0)}_{ik} - \frac{1}{2} R_{ik} F_{mn}F^{mn} -
\frac{1}{2} g_{ik} \nabla^l \nabla_l (F_{mn}F^{mn}) + \frac{1}{2}
\nabla_{i} \nabla_{k} (F_{mn}F^{mn})  \,, \label{part1}
\end{align}
\begin{align}
T^{(2)}_{ik} &= - \frac{1}{2}g_{ik}\left[ \nabla_{m}
\nabla_{l}(F^{mn}F^{l}_{\ n} ) - R_{lm}F^{mn}F^{l}_{\ n}\right]
-F^{ln}(R_{il}F_{kn} + R_{kl}F_{in}) - R^{mn} F_{im} F_{kn} \nn &-
\frac{1}{2} \nabla^l \nabla_l (F_{in}F_{k}^{\ n}) +
\frac{1}{2}\nabla_l \left[ \nabla_i(F_{kn}F^{ln}) +
\nabla_k(F_{in}F^{ln}) \right] \,, \label{part2}
\end{align}
\begin{align}
T^{(3)}_{ik} = \frac{1}{4}g_{ik} R^{mnls}F_{mn}F_{ls} -
\frac{3}{4}F^{ls}(F_{i}^{\ n}R_{knls}+F_{k}^{\ n}R_{inls}) -
\frac{1}{2}\nabla_{m} \nabla_{n}(F_{i}^{\ n}F_{k}^{\ m} + F_{k}^{\
n}F_{i}^{\ m})\,. \label{part3}
\end{align}
We are happy to stress that with the ansatz (\ref{6}) for the metric
(\ref{3}) the system of equations (\ref{standardform}) transforms
into one equation
\begin{equation}
r N^{\prime} \left(1+\frac{\kappa q}{r^4}\right) + N \left(1 - 3q
\frac{\kappa}{r^4} \right) = 1 - \frac{\kappa}{2r^2} - 3 q
\frac{\kappa}{r^4} \,, \label{N}
\end{equation}
whose solution is
\begin{equation}
N(r)=1+\frac{r^2(\kappa -4Mr)}{2(r^4+\kappa q)} \,. \label{N1}
\end{equation}
Here $\kappa$ is the convenient constant, $\kappa=8\pi\mu^2$. This
solution is regular in the center ($N(0)=1$) and satisfies the
asymptotic condition $N(\infty)=1$. If the mass $M$ is less than
its critical value $M_{({\rm crit})}$, where
\begin{align}
M_{{(\rm
crit})}=\frac{r_{\!*}}{6}\left(4+\frac{\kappa}{r_{\!*}^2}\right)\,,
\quad
r_{\!*}=\frac{\sqrt{\kappa}}{2}\sqrt{\left(\sqrt{1+\frac{48\,q}{\kappa}}+1\right)}\,,
\label{rstar}
\end{align}
the metric (\ref{3}) with (\ref{N1}) has no horizons as in
\cite{BaZa07}.

\section{Electrodynamic description of the photon propagation}

Linear electrodynamics allows us to consider the dynamics of test
photons in terms of microscopic field strength $f_{ik}$ and
induction $h_{ik}$, which satisfy the equations
\begin{equation}
\nabla_k h^{ik} = 0 \,, \quad h^{ik}=C^{ikmn} f_{mn} \,, \quad
\nabla_k f^{*ik} = 0 \,. \label{41}
\end{equation}
Clearly, the microscopic tensor $f_{mn}$ is the analog of the
macroscopic Maxwell tensor $F_{mn}$, describing the background
electromagnetic field, and $h^{ik}$ is the analog of the tensor
$H^{ik}$ (compare (\ref{4}), (\ref{40}) and (\ref{41})). We have to
stress that the tensor of material coefficients $C^{ikmn}$ in the
formula (\ref{41}) describing the photon propagation, is the same
one as in (\ref{40}) for the background field. Thus, the non-minimal
interaction influences the photon by two ways: first, via the
constitutive equations (see (\ref{40}), (\ref{2})), second, via the
gravitational field, whose potentials (\ref{N1}) depends on the
parameter of non-minimal coupling $q$.

\subsection{Associated metrics}

The tensor $C^{ikmn}$ can be represented algebraically as a
quadratic combination of the so-called {\it associated} metrics
$g^{im (\alpha)}$  \cite{HehlObukhov,BZ05}
\begin{equation} C^{ikmn} {=} \frac{1}{4 \hat{\mu}}
\sum_{(\alpha)(\beta)}G_{(\alpha)(\beta)}\left[g^{im (\alpha)} \
g^{kn (\beta)} {-} g^{in (\alpha)} \ g^{km (\beta)} \right],
\label{supergeneral}
\end{equation}
where $G_{(\alpha)(\beta)}$ and $\hat{\mu}$ compose a set of
parameters, introduced in \cite{BZ05}. When the medium is spatially
isotropic, this decomposition requires one associated metric; in the
case of uniaxial spatial symmetry one needs two associated metrics
\cite{Perlick,BZ05}. In the framework of our model vacuum
interacting with curvature can be regarded as an uniaxial
quasi-medium, where the radial direction is the selected one.
Indeed, let us use the standard definitions of the dielectric
permittivity tensor $\varepsilon^{ik}$, magnetic impermeability
tensor $(\mu^{-1})_{pq}$, and tensor of magnetoelectric coefficients
$\nu_p^{\ m}$, given by \cite{HehlObukhov,EM}
\begin{gather}
\varepsilon^{im} = 2 U_{k} U_{n} C^{ikmn} \,, \quad (\mu^{-1})_{pq}
= -\frac{1}{2} \eta_{pik} C^{ikmn} \eta_{mnq} \,, \quad \nu_p^{\ m}
= \eta_{pik} C^{ikmn} U_n \,. \label{main}
\end{gather}
Here $U^k=\delta^k_t / \sqrt{N}$ is the velocity four-vector,
associated with the magnetic charge (at rest), $\eta_{pik} =
\epsilon_{pikm}U^m$ and $\epsilon_{pikm}$ is the Levi-Civita tensor.
In the spherically symmetric model under consideration the tensor
$\nu_p^{\ m}$ is equal to zero identically, other tensors have the
following non-vanishing components:
\begin{equation}\label{Bmu}
\varepsilon_{||} \equiv \varepsilon^{r}_{r} = 2 C^{rt}_{\ \ rt} = 1
+ 2 {\cal R}^{r t}_{ \ \ r t}  \,,
\end{equation}
\begin{equation}\label{Bepsilon}
\varepsilon_{\bot} \equiv \varepsilon^{\theta}_{\theta} =
\varepsilon^{\varphi}_{\varphi}
 = 1 + 2 {\cal R}^{\theta t}_{ \ \ \theta t}  =
 1 + 2 {\cal R}^{\varphi t}_{ \ \ \varphi t} \,,
\end{equation}
\begin{equation}\label{Beps}
\frac{1}{\mu_{||}} \equiv (\mu^{-1})^{r}_{r} = 1 + 2 {\cal
R}^{\theta \varphi}_{ \ \ \theta \varphi} \,,
\end{equation}
\begin{equation}\label{Beps1}
\frac{1}{\mu_{\bot}} \equiv (\mu^{-1})^{\theta}_{\theta} =
(\mu^{-1})^{\varphi}_{\varphi} = 1 + 2 {\cal R}^{\theta r}_{ \ \
\theta r} = 1 + 2 {\cal R}^{\varphi r}_{ \ \ \varphi r}  \,.
\end{equation}
For the metric (\ref{3}) with (\ref{N1}) we obtain
\begin{align}
{\cal R}^{r \theta}_{\ \ r \theta} = {\cal R}^{r \varphi}_{\ \ r
\varphi} = {\cal R}^{\theta t}_{\ \ \theta t} = {\cal R}^{\varphi
t}_{\ \ \varphi t} =  - \frac{r}{(r^4 + a^4)^3}[6q Mr^8 - 3a^4 r^7 -
24 q M a^4 r^4 +5 a^8 r^3 + 2q M a^7 ] \,, \label{as4}
\end{align}
\begin{align}
{\cal R}^{r t}_{\ \ r t} =  \frac{r}{(r^4+a^4)^3}[12q M r^8 - 7 a^4
r^7 - 76 q M a^4 r^4 +17 a^8 r^3 + 8 q M a^8 ] \,, \label{as5}
\end{align}
\begin{align}
{\cal R}^{\theta \varphi}_{\ \ \theta \varphi} =
\frac{r^4}{(r^4+a^4)^3}[12q M r^5
 - 5 a^4 r^4 -20 q M a^4 r + 3 a^8 ] \,. \label{as6}
\end{align}
A new parameter $a$ with dimensionality of length is defined as
follows:  $a^4 = \kappa q$. Since $\varepsilon^{\theta}_{\theta} =
\varepsilon^{\varphi}_{\varphi}$ and $(\mu^{-1})^{\theta}_{\theta}
= (\mu^{-1})^{\varphi}_{\varphi}$, the medium can be considered as
uniaxial. Moreover, due to the equality ${\cal R}^{r \theta}_{\ \
r \theta} =  {\cal R}^{\theta t}_{\ \ \theta t}$ the relation
$\varepsilon_{\bot} \mu_{\bot} =1$ takes place. The tensor
$C^{ikmn}$ can be reconstructed algebraically as
\begin{align}
C^{ikmn} = \frac{1}{2 \hat{\mu}} \biggl\{ \left[g^{im (A)} g^{kn
(A)} - g^{in (A)} g^{km (A)} \right] &- \gamma \biggl[\left(g^{im
(A)}{-}g^{im (B)} \right)\left(g^{kn (A)}{-}g^{kn (B)}\right) \nn &-
\left(g^{in (A)}{-}g^{in (B)}\right)\left(g^{km (A)}{-}g^{km
(B)}\right) \biggr] \biggr\} \,, \label{mainlong1}
\end{align}
where the associated metrics $g^{ik (A)}$ and $g^{ik (B)}$ are
\begin{equation}
g^{im (A)} = U^{i} U^{k} +
\frac{\varepsilon_{\bot}}{\varepsilon_{||}} \Delta^{im} +
\frac{(\varepsilon_{\bot}-\varepsilon_{||})}{\varepsilon_{||}}
X^{i}_{(r)}X^{m}_{(r)} \,, \label{mainlong2}
\end{equation}
\begin{equation}
g^{im (B)} = U^{i} U^{k} + \frac{\mu_{\bot}}{ \mu_{||}} \Delta^{im}
+ \frac{(\mu_{\bot}-\mu_{||})}{\mu_{||}} X^{i}_{(r)}X^{m}_{(r)} \,.
\label{mainlong3}
\end{equation}
Here $X^k_{(r)} = \delta^k_r \sqrt{N(r)}$ is the radial tetrad
space-like four-vector and $\Delta^k_m \equiv \delta^k_m - U^kU_m $
is the projector. The representation (\ref{mainlong1}) is a
particular case of the decomposition (\ref{supergeneral}) with
\[
G_{(A)(A)}+ G_{(B)(A)} = 1 \,, \quad G_{(A)(B)}+G_{(B)(B)} = 0 \,,
\quad G_{(A)(B)} = G_{(B)(A)} = \gamma \,,
\]
\begin{equation}\label{Bg111}
\frac{1}{\hat{\mu}} = \varepsilon_{||} \,, \quad \frac{1}{\gamma} =
1 - \frac{\varepsilon_{||}\mu_{\bot}}{\varepsilon_{\bot}\mu_{||}}
\,.
\end{equation}
Mention that $g^{ik (B)}$ can be obtained from $g^{ik (A)}$ by a
formal replacement of the symbol $\varepsilon$ by the symbol $\mu$.

\subsection{Optical metrics}

In the approximation of geometrical optics \cite{Synge} the
potential four-vector $A_l$ and the field strength tensor can be
represented as
\begin{align}
A_l = a_l e^{i \Phi} \,, \quad F_{mn} = \nabla_{m}A_{n} {-}
\nabla_{n}A_{m}= i\left(k_{m}A_{n} {-} k_{n}A_{m}\right), \label{F}
\end{align}
respectively, where $\Phi$ is the phase, $a_l$ is a slowly varying
amplitude, and $k_m$ is a wave four-vector:
\begin{equation}
k_m \equiv \nabla_m \Phi \,. \label{A}
\end{equation}
In the leading-order approximation the Maxwell equations reduce to
\begin{equation}
 C^{ilmn} k_l k_m  a_n  = 0 \,. \label{Maxgo}
\end{equation}
This system of linear equations with respect to $a_n$ admits
non-trivial solutions, when four components of $k_l$ satisfy the
Fresnel equation (see, e.g., \cite{HehlObukhov} for details), which
is usually called the dispersion equation. Using the decomposition
(\ref{mainlong1}) with (\ref{mainlong2}) and (\ref{mainlong3}) in
(\ref{Maxgo}) after some algebra we obtain the dispersion equations
for two different principal cases.

{\it (i)}. The first case relates to the propagation of
electromagnetic wave with non-vanishing longitudinal polarization,
i.e., with $a_{r} \neq 0$. The Fresnel equation admits the solution
\begin{equation}
\frac{\omega^2}{c^2}  +  k_r k^r + \frac{1}{\varepsilon_{||}
\mu_{\bot}}(k_{\theta} k^{\theta}+k_{\varphi}k^{\varphi}) = 0 \,,
\label{M1}
\end{equation}
where $\omega \equiv U^i k_i$ is a frequency. This dispersion
relation can be reconstructed as an eikonal-type equation
\begin{equation}
g^{im (A)} k_i k_m = 0 \,, \label{M3}
\end{equation}
where $g^{im (A)}$ is the first associated metric (\ref{mainlong2}).

{\it (ii)}. The second case is characterized by $a_{r} = 0$. The
Fresnel equation admits the solution
\begin{equation}
\frac{\omega^2}{c^2}  +  k_r k^r + \frac{1}{\varepsilon_{\bot}
\mu_{||}}(k_{\theta} k^{\theta}+k_{\varphi}k^{\varphi}) = 0 \,,
\label{M2}
\end{equation}
which can be rewritten as
\begin{equation}
g^{im (B)} k_i k_m = 0 \,, \label{M4}
\end{equation}
where $g^{im (B)}$ is the second associated metric
(\ref{mainlong3}).

When the electromagnetic wave propagates  along the radial
direction, i.e., when $k_{\theta}=k_{\varphi}=0$ and $k_rk^r =-k^2$,
the dispersion relations (\ref{M1}) and (\ref{M2}) give the same
result, $\omega^2 = k^2c^2$, providing the phase velocity to be
equal to speed of light in standard vacuum. The velocity of
propagation in the transversal directions (i.e., when $k_r=0$)
depends on the polarization: when $a_r \neq 0$  the phase velocity
of the wave is $V_{(1)}=
\sqrt{\frac{\varepsilon_{\bot}}{\varepsilon_{||}}}$, when $a_r=0$ it
is $V_{(2)}= \sqrt{\frac{\mu_{\bot}}{\mu_{||}}}$.

\section{Photon dynamics}

The associated metrics obtained in Subsection 3.1 are the optical
ones, i.e., they satisfy the eikonal equation. This means that the
motion of a photon influenced by the non-minimal interaction can be
described by null geodesic lines in the effective space-times with
optical metrics  $g^{im (A)}$ or $g^{im (B)}$, depending on
polarization. Three metrics: the metric of space-time (\ref{3}), the
first and second optical metrics (\ref{mainlong2}) and
(\ref{mainlong3}), can be written in general form
\begin{align}
ds^2_{(E)} = N(r) dt^2 - \frac{dr^2}{N(r)} - r^2 Y_{(E)}(r)
(d\theta^2 + \sin^2\theta d\varphi^2)\,, \label{M12}
\end{align}
where the index $E$ takes the values $0,\ A,\ B$ and
$Y_{(0)}(r){=}1$. For the sake of simplicity we consider below the
monopole with $M=0$. It is characterized by three interesting
features: first, $N(r) \geq 1$ for arbitrary $r$, thus, $N(r)$ does
not reach zero; second, $N(r)$ reaches its maximum at $r=a$, the
maximal value being $N_{(max)} = 1+ \sqrt{\frac{\kappa}{16q}}$;
third, $N(0)=1$ and curvature invariants are regular in the center
$r=0$. For this particular model we have
\begin{equation}
Y_{(A)}(r) = \frac{1-11 \xi + 37 \xi^2 + \xi^3}{1+9 \xi - 7 \xi^2 +
\xi^3} \,, \label{M13}
\end{equation}
\begin{equation}
Y_{(B)}(r) = \frac{1 + 9 \xi - 7 \xi^2 + \xi^3}{1 - 7 \xi +9 \xi^2 +
\xi^3} \,, \label{M14}
\end{equation}
where $\xi$ is dimensionless positive quantity $\xi=\kappa q /
r^4$. The polynomial $1-11 \xi + 37 \xi^2 + \xi^3$ has no real
positive zeros, i.e., $Y_{(A)}(r)\neq 0$. The equalities
$Y_{(A)}(r) = \infty$ and simultaneously $Y_{(B)}(r)=0$ take
place, when $1 + 9 \xi -7 \xi^2 + \xi^3=0$, i.e., at $\xi = \xi_1
\approx 1.85$ and $\xi = \xi_2 \approx 5.25$. Finally, $Y_{(B)}(r)
= \infty$, when $\xi = \xi_3 \approx 0.19$ and $\xi =\xi_4 \approx
0.54$. Thus, despite the metric $ds^2_{(0)}$ is regular
everywhere, the first and second optical metrics have singular
points, when the angular functions $Y_{(A)}(r)$ and $Y_{(B)}(r)$
vanish. Physical interpretation of such (dynamic) singularities
will be done in the next Subsection in the course of description
of radial, tangent and inclined particle motion.

\subsection{Photon trajectories}
\EWFigure{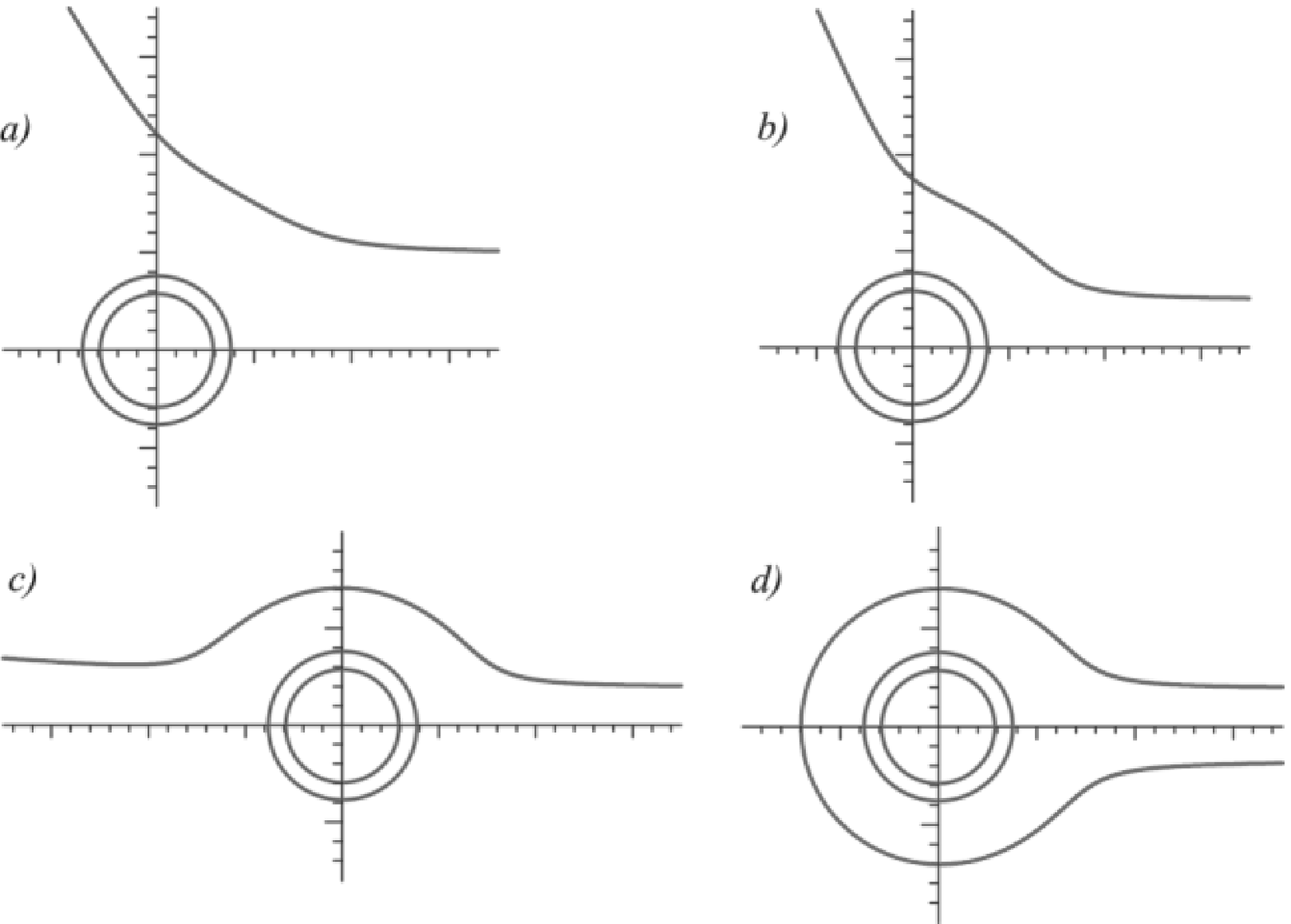}{Rays for optical A-metric at the impact
parameter $J>J^{(A)}_{\rm crit}\approx 0.794659$: (a) $J=2$, (b)
$J=1$, (c) $J=0.8$, (d) $J=0.79466$.} \EWFigure{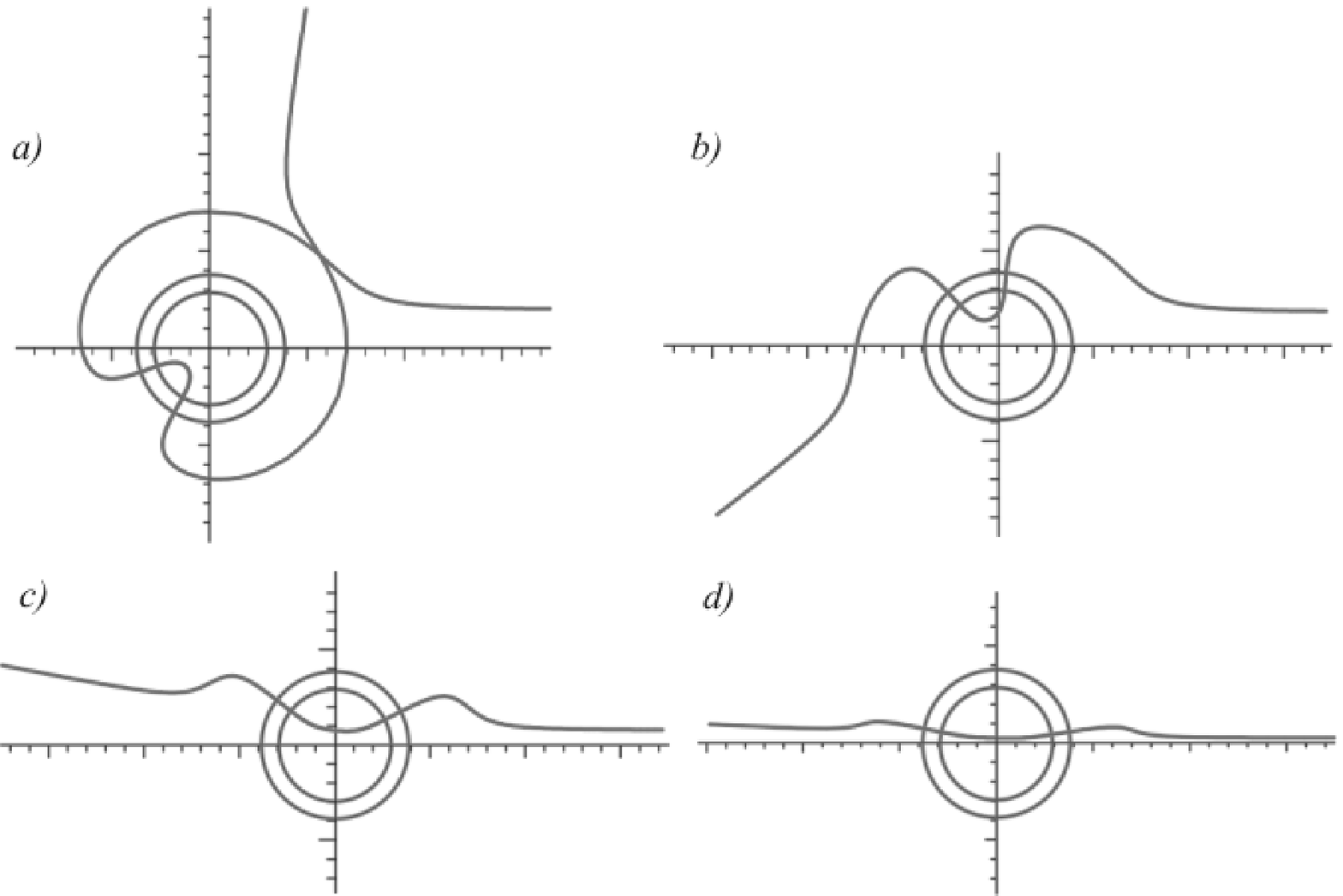}{Rays for
optical A-metric at the impact parameter $J<J^{(A)}_{\rm
crit}\approx 0.794659$: (a) $J=0.7941$, (b) $J=0.7$, (c) $J=0.3$,
(d) $J=0.1$.} \EWFigure{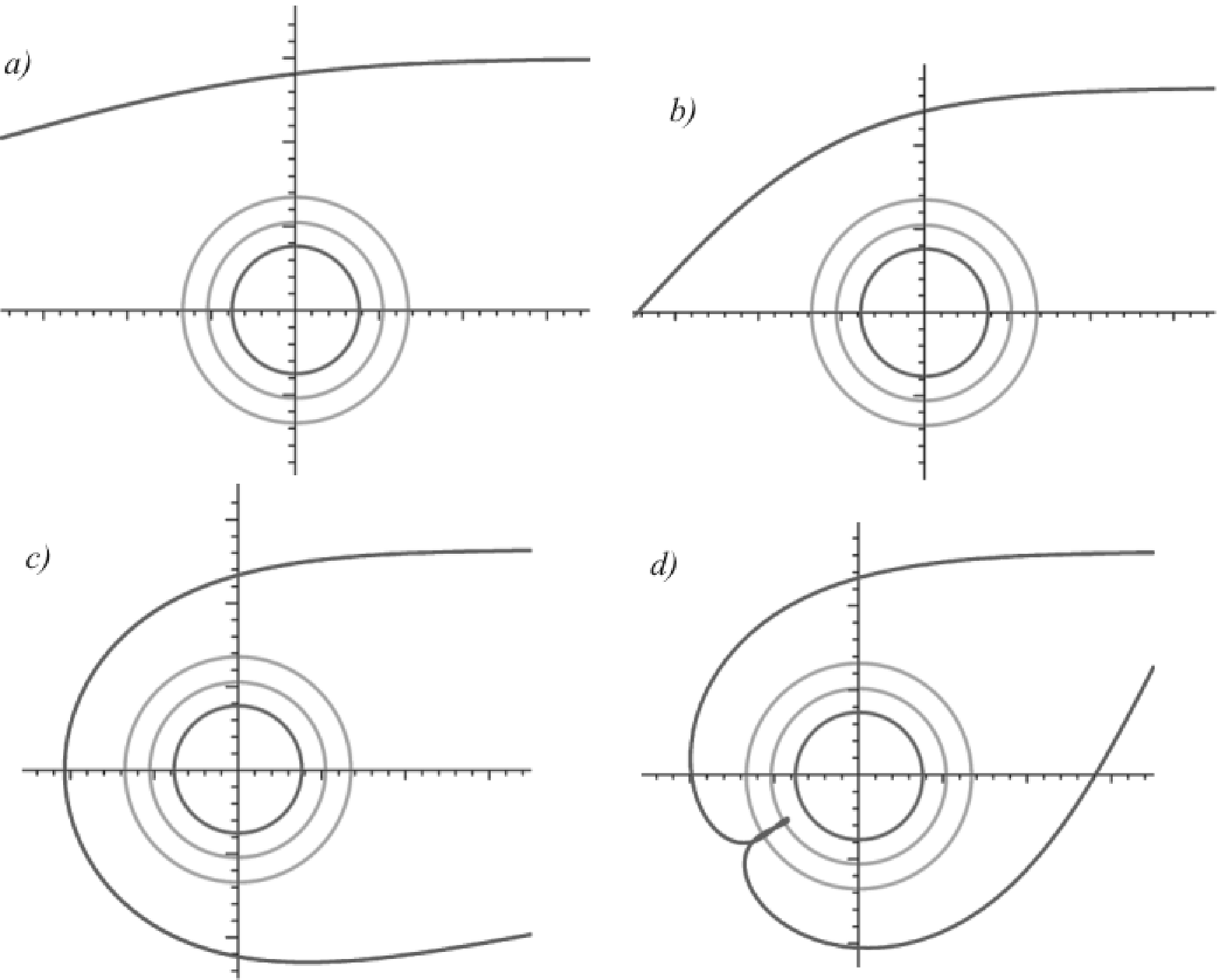}{Rays for optical B-metric at the
impact parameter $J>J^{(B)}_{\rm crit}\approx 5.29385$: (a) $J=6$,
(b) $J=5.4$, (c) $J=5.294$, and (d) $J<J^{(B)}_{\rm crit}$,
$J=5.29$.} \EWFigure{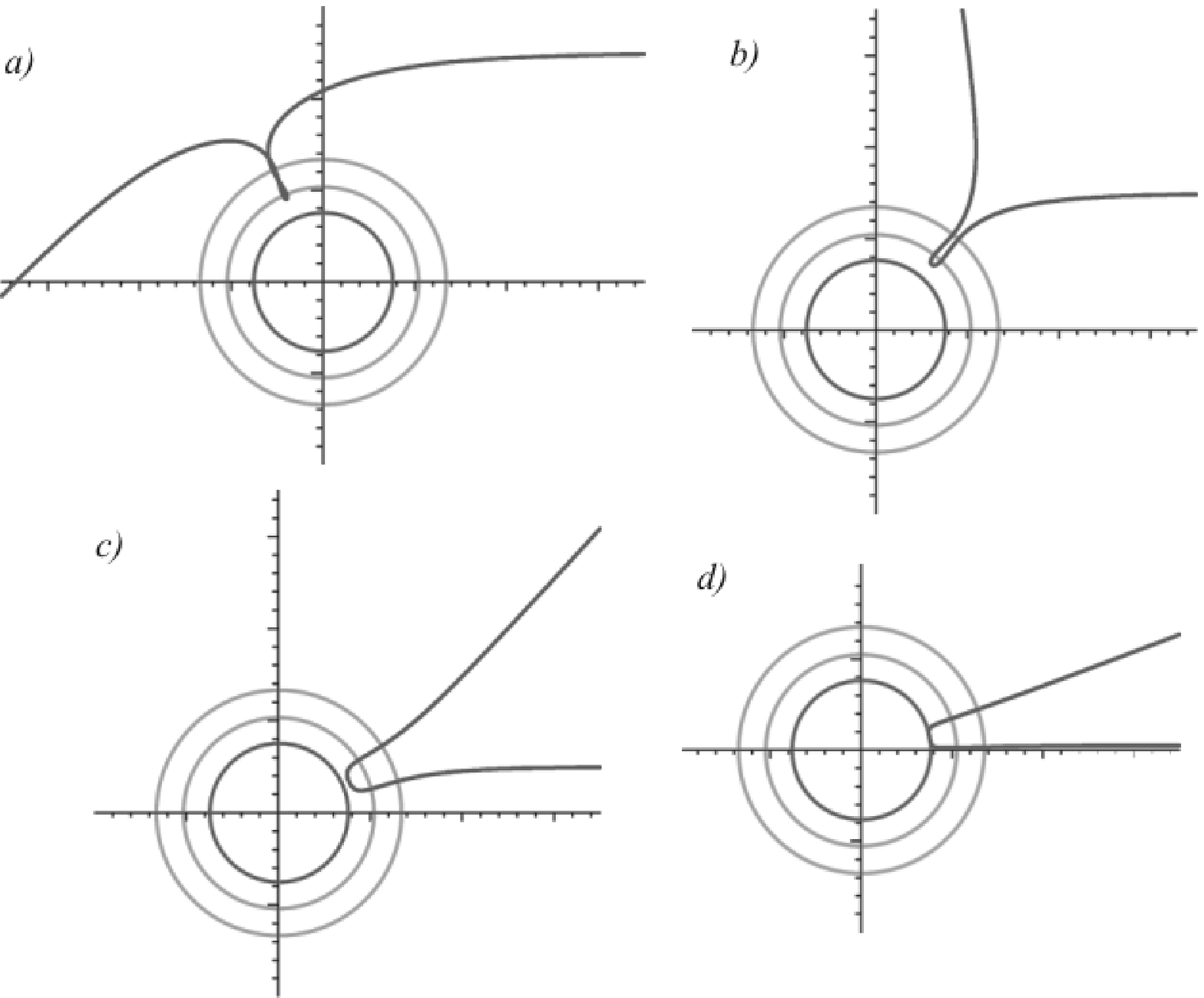}{Rays for optical B-metric at the
impact parameter $J<J^{(B)}_{\rm crit}\approx 5.29385$: (a) $J=5$,
(b) $J=3$, (c) $J=1$, (d) $J=0.1$.}

Let us consider null geodesic lines in the effective space-times
with optical metrics (\ref{mainlong2}) and (\ref{mainlong3}). The
geodesic equations have the standard form
\begin{equation}
\frac{d^2 x^k }{d\tau^2} + \Gamma^k_{jl (E)} \frac{dx^j}{d\tau}
\frac{dx^l}{d\tau} = 0  \,, \label{dyna1}
\end{equation}
where the Christoffel symbols $\Gamma^k_{jl (E)}$, $E=0,\ A,\ B$ are
constructed on the base of the corresponding metric. As in classical
case (see, e.g., \cite{Weinberg}) the photon moves in the plane, and
one can choose this plane as the equatorial one, $\theta = \pi/2$.
Other integrals have the form
\begin{equation}
\frac{d\varphi}{d\tau} = \frac{J}{r^2 Y_{(E)}(r)} \,, \quad
\frac{dt}{d\tau} = \frac{1}{N(r)} \,, \label{time1}
\end{equation}
\begin{equation}
\left(\frac{dr}{d\tau}\right)^2 =  1 - \frac{J^2 N(r)}{r^2
Y_{(E)}(r)} \,, \label{rad11}
\end{equation}
providing the relation
\begin{equation}
ds^2_{(E)} = \left[g_{(E) mn } \frac{dx^m}{d\tau}
\frac{dx^n}{d\tau} \right] d\tau^2 =0 \,, \label{time2}
\end{equation}
where the quantity $J$ is the impact parameter. As a consequence of
(\ref{time1}) and (\ref{rad11}) the following relation takes place:
\begin{equation}
\left(\frac{dr}{d\varphi}\right)^2 = r^2 Y_{(E)}(r) \left[
\frac{1}{J^2} r^2 Y_{(E)}(r)- N(r) \right] \,. \label{time23}
\end{equation}
The formulas (\ref{time1})-(\ref{time23}) are the non-minimally
modified formulas for the well-known integrals of motion
\cite{Weinberg} (they coincide with them, when $Y_{(E)}(r)=1$ only).
Let us emphasize seven interesting details of these modified
integrals of motion.

(I) Radial motion is non-sensitive to the non-minimal coupling.
Indeed, when $\theta = \frac{\pi}{2}$ and $\varphi = const$, the
parameter $J$ vanishes and the relation $\frac{dr}{d\tau} = \pm 1$
is valid for each $E=0,A,B$. In this case the velocity of photon
coincides with speed of light in standard vacuum.

(II) When the motion is not pure radial (inclined), the condition
$\frac{dr}{d\varphi}=0$ describes the extrema of the curve
$r(\varphi)$, giving the distance of the closest approach. When the
curvature interaction is absent, i.e., $q=0$, we obtain from
(\ref{time23}) the standard condition for the minimal distance
$r_0$: $r^2_0 - J^2 N(r_0)=0$. In the non-minimal model $Y_{(A)}$
and $Y_{(B)}$ are the functions of radius (see (\ref{M13}) and
(\ref{M14})), and the set of solutions to the equation
$Y_{(E)}(r^2Y_{(E)}-J^2 N)=0$ is much more sophisticated than in the
minimal model.

(III) Formally speaking, the optical metrics become non-Lorentzian,
i.e., are of the signature $+-++$, when $Y_{(E)}(r)$ is negative,
and are singular at the points, where $Y_{(E)}(r)=0$. From the
dynamic point of view the negative values of $Y_{(E)}(r)$ are,
nevertheless, admissible. Indeed, the angular integral of motion in
(\ref{time1}) remains consistent, but the direction of rotation
becomes opposite; the right-hand-sides of the equations
(\ref{rad11}) and (\ref{time23}) remain positive, as it should be.

(IV) The functions $Y_{(E)}(r)$ change the sign at the points, where
$Y_{(E)}(r)=0$ or $Y_{(E)}(r)= \infty$. The first possibility can be
realized for the B-ray only, when $\xi = \kappa q / r^4$ takes the
values 1.85 or 5.25. In this case the equation (\ref{time23}) gives
the condition $\frac{dr}{d\varphi}=0$, indicating the point of
extremum for the curve $r(\varphi)$ (see item III). The conditions
$Y_{(E)}(r)=\infty$ can be realized at $\xi=1.85$ and $\xi = 5.25$
for the A-ray, and at $\xi=0.19$ and $\xi = 0.54$ for the B-ray. At
these points the angular frequency $\frac{d \varphi}{d \tau}$
vanishes (see (\ref{time1})) and the radial velocity $\frac{dr}{d
\tau}$ coincides with speed of light in the standard vacuum (see
(\ref{rad11})). Since the relation $\frac{dr}{d\varphi}=\infty$
holds at these points, the rays are directed radially.

(V) The relation $\frac{dr}{d \tau}=0$ is self-contradictory, when
$Y_{(E)}$ is negative (see (\ref{rad11})). This means that tangent
waves can not propagate, and we can indicate this zone as
inaccessible one for them. For the tangent A-ray, there is only one
inaccessible zone at $1.85< \xi< 5.25$, for the tangent B-ray the
second one appears at $0.19< \xi< 0.54$. The corresponding boundary
spheres can be indicated as trapped surfaces (see, e.g.,
\cite{AG1,AG2} for details).

(VI) For A- and B-rays there are specific ``critical'' values of the
impact parameters \[J_{\rm
crit}^{(A),(B)}=\max\left(\frac{N}{r^2Y_{(A),(B)}}\right)\,.\] If
$J>J_{\rm crit}$, the corresponding rays do not reach the surface
with $Y_{(E)}(r)=\infty$. Mention that in minimal case such
``critical'' values of the impact parameters do not exist.

(VII) Since the space-time metric of the non-minimal monopole is
regular, i.e., $N(r) \neq 0$ and $N(r) \neq \infty$, in the $t-r$
cross-section there is no singularities.

\subsection{Numerical modeling of the photon trajectories}

Pictures presented in the Fig. 1-4 illustrate the dependence
$r(\varphi)$ for rays, which are null geodesics of the optical
metrics $g^{ik}_{(A)}$ and $g^{ik}_{(B)}$. The rays start from
infinity, the impact parameter being equal to the height of the
starting point of curve. We put $\kappa=1$, $q=10$. In the Fig. 1, 2
two circles relate to $r=R_1\equiv\sqrt[4]{\kappa q/1.85}$ and
$r=R_2\equiv\sqrt[4]{\kappa q/5.25}$; in the Fig. 3, 4 three circles
relate to $r=R_3\equiv\sqrt[4]{\kappa q/0.19}$,
$r=R_4\equiv\sqrt[4]{\kappa q/0.54}$, and $r=R_1$, respectively. The
picture (a) in the Fig. 2 contains one self-intersection point, the
picture (d) in the Fig. 3 and the picture (a) in the Fig. 4 contain
two self-intersection points. The region inside the circle $r=R_1$
is inaccessible zone for the inclined B-rays. These figures show
that the non-minimal Dirac monopole can be considered as a
scattering center for inclined rays. Scattering laws for A- and
B-rays are different confirming the birefringent character of
non-minimal interaction.

\section{Conclusions}

Qualitative and numerical analysis of the photon orbits in the
vicinity of non-minimal Dirac monopole with regular metric has
demonstrated the following interesting features.

1. Propagation of the electromagnetic waves in the vicinity of
non-minimal Dirac monopole is characterized by birefringence,
induced by curvature, i.e., the phase velocities of waves depend on
their polarization. Two different optical metrics should be
introduced to describe two principal states of polarization.

2. The metric of the non-minimal Dirac monopole, obtained and
discussed in this paper, is regular, thus, all the singularities of
the optical metrics have a dynamic origin and are supported by the
non-minimal (curvature induced) interaction of the gravitational and
electromagnetic fields.

3. The points of self-intersection, the points of the closest
approach, the reverse points, etc. in the photon trajectories can be
recognized and catalogued for different combinations of the values
of the impact parameter and the parameter of non-minimal coupling.
We discussed here only the principal pictures.

\Acknow This work was partially supported by the DFG through project
No. 436RUS113/487/0-5. The authors are grateful to Claus
L\"ammerzahl for valuable remarks.


\small
\begin{thebibliography}{99}

\bibitem{Dirac} P.A.M. Dirac, {\it Proc. R. Soc. Lond.} {\bf A
133}, 60 (1931).

\bibitem{000} F.A. Bais, R.J. Russell, \PRD 11 2692 (1975).

\bibitem{00} Y.M. Cho, P.G.O. Freund, \PRD 12 1588 (1975).

\bibitem{0} P.B. Yasskin, \PRD 12 2212 (1975).

\bibitem{1} P. Rossi, {\it Phys. Rep.} {\bf 86}, 317 (1982).

\bibitem{3} M.S. Volkov and D.V. Gal'tsov, {\it Phys. Rep.} {\bf 319}, 1 (1999).

\bibitem{2} K.A. Milton, {\it Rep. Prog. Phys.} {\bf 69}, 1637 (2006).

\bibitem{4}  Ya.M. Shnir, ``Magnetic Monopoles'', Springer-Verlag,
Berlin, 2005.

\bibitem{5} V. Kagramanova, J. Kunz, C. L\"ammerzahl, arxiv:
0708.1747 [gr-qc].

\bibitem{BaZa07} A.B. Balakin and A.E. Zayats, \PLB 644 294 (2007).

\bibitem{Horn} G.W. Horndeski, \JMP 19 668 (1978).

\bibitem{MHS} F. M\"uller-Hoissen and R. Sippel, \CQG 5 1473 (1988).

\bibitem{B1} A.B. Balakin, \CQG 14 2881 (1997).

\bibitem{BL05} A.B. Balakin and J.P.S. Lemos, \CQG 22 1867 (2005).

\bibitem{HehlObukhov} F.W. Hehl and Yu.N. Obukhov, ``Foundations of Classical
Electrodynamics: Charge, Flux, and Metric'', Birkha\"user, Boston,
2003.

\bibitem{Perlick} V. Perlick, ``Ray Optics, Fermat's Principle, and
Applications to General Relativity'', Springer-Verlag, Berlin, 2000.

\bibitem{BZ05} A.B. Balakin and W. Zimdahl, \GRG 37 1731 (2005).

\bibitem{AG1} C.Barcel\'o, S. Liberati, and M. Visser, {\it Living Rev. Rel.}
{\bf 8}, 12 (2005).

\bibitem{AG2} M. Novello, M. Visser, and G. Volovik (Eds.), ``Artificial
Black Holes'', World Scientific, Singapore, 2002.

\bibitem{EM} A.C. Eringen and G.A. Maugin, ``Electrodynamics of
Continua'', Springer-Verlag, New York, 1989.

\bibitem{Synge} J.L. Synge, ``Relativity: The General
Theory'', North-Holland, Amsterdam, 1971.

\bibitem{Weinberg} S. Weinberg, ``Gravitation and Cosmology'', Wiley, New York, 1972.



\end{thebibliography}
\end{document}